\documentclass[preprint,onecolumn,superscriptaddress,tightenlines,nofootinbib,11pt]{article}
\pdfoutput=1 
\usepackage{graphicx}
\usepackage{epsfig}
\usepackage{enumitem}
\usepackage{amsmath,latexsym,amssymb,hyperref,enumerate}
\usepackage{amsfonts}
\usepackage{url}
\usepackage{color}
\usepackage{cancel}
\usepackage{cite}
\usepackage[font={sf}]{caption}
\hypersetup{colorlinks,citecolor= nicegreen,linkcolor= nicered}
\definecolor{nicered}{rgb}{0.7,0.1,0.1}
\definecolor{nicegreen}{rgb}{0.1,0.5,0.1}

\renewcommand{\thefootnote}{\fnsymbol{footnote}}

\begin{document}

\vspace*{1.0cm}

\centerline{\bf\Large\textcolor[rgb]{0,0,0}{A Reappraisal on Dark Matter Co-annihilating}\vspace*{0.2cm}}
\centerline{\bf\Large\textcolor[rgb]{0,0,0}{with a Top/Bottom Partner}}
\vspace*{1.2cm}

\centerline{\bf 
{Wai-Yee~Keung\,$^1$\,\footnote{\tt keung@uic.edu}}, \ \ 
{Ian~Low\,$^{2,3}$\,\footnote{\tt ilow@northwestern.edu}}, \ \  
{Yue~Zhang\,$^2$\,\footnote{\tt yuezhang@northwestern.edu}}
}

\vspace*{1.0cm}

\centerline{\em $^1$Physics Department, University of Illinois at Chicago, IL 60607, USA\vspace*{0.1cm}}
\centerline{\em $^2$Department of Physics and Astronomy, Northwestern University, Evanston, IL 60208, USA\vspace*{0.1cm}}
\centerline{\em $^3$High Energy Physics Division, Argonne National Laboratory, Argonne, IL 60439, USA}

\vspace{1.5cm}

\parbox{16.5cm}{
{\sc Abstract:} 
We revisit the calculation of relic density of dark matter particles co-annihilating with a top or bottom partner, by properly including the QCD bound-states (onia) effects of the colored partners, as well as the relevant electroweak processes which become important in the low mass region. We carefully set up the complete framework that incorporates the relevant contributions and investigate their effects on the cosmologically preferred mass spectrum, which turn out to be comparable in size to those coming from the Sommerfeld enhancement. We apply the calculation to three scenarios: bino-stop and bino-sbottom co-annihilations in supersymmetry, and a vector dark matter co-annihilating with a fermionic top partner. In addition, we confront our analysis of the relic abundance with  recent direct detection experiments and collider searches at the LHC, which have important implications in the bino-stop and bino-sbottom scenarios. In particular, in the bino-stop case recent LHC limits have excluded regions of parameter with a direct detection rate that is above the neutrino floor. }

\vspace{1.5cm}
NUHEP-TH/17-02

\newpage

%%%%%%%%%%%%%%%%%%%%%%%%%%%%%%%%%%%%%%%%%%%%%%%
\section{Introduction}
%%%%%%%%%%%%%%%%%%%%%%%%%%%%%%%%%%%%%%%%%%%%%%%

Understanding nature of dark matter (DM) in our universe is of tremendous importance and interest to particle physics and cosmology. If the DM is made of a new particle, a new theory beyond the standard model (SM) will be called for. 

This work is devoted to studying a scenario where the DM particle $X$ is accompanied in the mass spectrum with a heavier partner $Y$ which is colored and participates in the strong interaction. Both are odd under a $Z_2$ symmetry which is good enough so that the DM is cosmologically stable. Their quantum numbers are chosen such that a renormalizable interaction is allowed among the DM, its partner and a third generation SM quark. We will focus on the class of models where the DM is a gauge singlet under the SM so that if its partner is too heavy and decoupled it has no efficient annihilation channels and will be thermally overproduced. If the partner mass is nearby, the DM could obtain the observed relic abundance in the early universe through the so-called co-annihilation mechanism~\cite{Griest:1990kh}. During the thermal freeze out the DM particles are most efficiently depleted by converting into the partner particles which can annihilate away through strong interactions. By adjusting the DM-partner mass difference (typically less than $\sim$\,100\,GeV), the correct relic abundance can be explained for a wide range of DM mass from hundreds of GeV to multiple TeV scale. The co-annihilation picture can be implemented in many well motivated contexts of new physics such as supersymmetry, or extra dimension theories. Phenomenologically, such a scenario can be interesting because the current LHC constraint on the color particle mass is weaker (much less than a TeV) with a compressed spectrum.

The aim of our work is to carry out a precision calculation of the DM relic abundance via co-annihilation. To this end, the following three effects must be taken into account.

\begin{itemize}[rightmargin=\dimexpr\linewidth-16cm-\leftmargin\relax]

\item First, in the annihilation of the colored co-annihilating partner $Y$ and its antiparticle, the gluon exchange between the two initial state colored particles can strongly distort their wavefunction at the origin giving rise to the Sommerfeld effect~\cite{Hisano:2006nn, Cirelli:2007xd, ArkaniHamed:2008qn}. During the thermal freeze out, the DM particles and their co-annihilating partners in the universe are no longer ultra-relativistic, therefore the Born level cross sections have to be modified to include such a non-perturbative effect. Such an effect has been included previously in~\cite{deSimone:2014pda, Harigaya:2014dwa, Ibarra:2015nca}. It is worth emphasizing that the Sommerfeld enhancement is in effect above the kinematic threshold of $2m_Y$, when there exists a long-range attractive force between two on-shell $Y$ particles. 

\item The second non-perturbative effect lies in the bound state formation channels where the co-annihilating partner $Y$ and its anti-particle radiatively captures each other and form a bound state $B$, before the annihilation takes place.  Such a bound state, the QCD onia, exists below the $2m_Y$ threshold,
\begin{equation}
m_B\  <\  2 m_Y \ ,
\end{equation}
which makes it distinct from the Sommerfeld enhancement that is active above $2m_Y$, between two on-shell $Y$ particles.
Such bound state effects have been found to play an important role in many aspects of DM interacting with a light mediator, from the production in the early universe~\cite{Wise:2014jva, vonHarling:2014kha, Ellis:2015vaa, Petraki:2015hla, Liew:2016hqo, ElHedri:2017nny, Mitridate:2017izz, Baldes:2017gzw}, to indirect detection~\cite{Pospelov:2008jd, MarchRussell:2008tu, Pearce:2013ola, An:2016gad, An:2016kie, Cirelli:2016rnw, Asadi:2016ybp, Johnson:2016sjs} and direct detection~\cite{Laha:2013gva, Wise:2014ola, Laha:2015yoa}, and to collider searches~\cite{Shepherd:2009sa, An:2015pva, Tsai:2015ugz, Bi:2016gca}. These motivate us to also include such an effect in the co-annihilation calculation. 

Based on the Kramers formula~\cite{kramers, Katkov}, generalizable to the QCD case, the most abundantly formed bound states have principle quantum number smaller than $\sim$\,$\alpha_S/v$, where $v$ is the DM relative velocity during the freeze out and numerically it is comparable to $\alpha_S$. As a result, the calculation including the ground state and first few excited states is expected to capture  the dominant contribution.

\item Third, although the mechanism of co-annihilation relies on the DM's converting into its partners, we also want to emphasize the importance of direct annihilation channels  of the DM at low temperature, which are often neglected in the previous simplified model analysis. These channels include the DM self-annihilation and the DM-partner co-annihilation channels. Once a DM-partner-quark coupling is introduced for efficient conversion, these channels are automatically induced, and their strength could be fixed in specific models (such as the simple SUSY models discussed below). During the freeze out, the rate of these channels could be competitive to or even more important than the strong interaction channels involving the DM partner. The key reason is that at low enough temperature  the DM number density is substantially higher than that of its partner due to the different Boltzmann suppression dictated by their mass difference.

We find that including the annihilation channels involving the DM tends to weaken the relative importance of non-perturbative effects. As a quantitative example, we find that in the bino-right-handed-sbottom co-annihilation case the bound state channel contribution to the overall annihilation rate can be as large as 30\%, while in the bino-right-handed-stop co-annihilation case it is always less than 10\% because right-handed stop has a larger hypercharge than sbottom.

\end{itemize}

In addition to the above refinements in the calculation of relic density, we also emphasize the interesting interplay with direct detection of DM and collider searches at the LHC. In particular, we point out that collider searches of stops and sbottoms in the regime of  "compressed spectra," where the stops and sbottoms are nearly degenerate with the DM particle, invoke the same underlying assumption as the bino-stop and bino-sbottom co-annihilations. As a result, limits from  LHC searches can directly constrain the parameter space in the co-annihilation scenarios. This is especially important considering the existence of a "neutrino floor" in direct detections, which sets a lower limit on the magnitude of direct detection rate that can be probed by conventional experimental setup. For example, we will demonstrate that, in the case of bino-stop co-annihilation, regions of parameter space with a direct detection rate above the neutrino floor is being excluded by the latest limits from the LHC Run 2.
%We will demonstrate that, for the bino-stop co-annihilation, the region of  parameter space with a direct detection rate below the neutrino floor is excluded by the latest collider searches for stops in the compressed spectrum. For the bino-sbottom co-annihilation, only the high mass region 

This paper is organized as the following. In section~\ref{secII}, we set up the general formalism for calculating the DM relic abundance in the co-annihilation mechanism with all the above effects included. We apply this formalism to calculate the required mass difference between DM and its partner in several simple models including the bino DM co-annihilating with a stop or sbottom in section~\ref{secIII}, and a vector boson DM co-annihilating with a fermionic top partner in section~\ref{secIV}. We confront our results with the existing and future experimental probes of the parameter space via DM direct detection and, in the case of bino-sbottom, the LHC searches.  We find including bound state channels could have strong implications in these searches. The conclusion is drawn in section~\ref{secV}.

During the course of this study, several recent works~\cite{Liew:2016hqo, Mitridate:2017izz, ElHedri:2017nny} appeared which partially overlap with our discussion on  the QCD bound state effects in DM co-annihilation. However, in~\cite{Liew:2016hqo,Mitridate:2017izz}, the annihilation channels directly involving the DM (the third bullet in the above) have been neglected in those studies. As will be shown, these neglected effects turn out to be important in the low temperature.
On the other hand, \cite{ElHedri:2017nny} considered the class of models where the DM-partner conversion happens through higher dimensional operators, which is a different scenario from the bino-stop, bino-sbottom, and vector DM and fermionic top partner co-annihilation models considered in this work.

%%%%%%%%%%%%%%%%%%%%%%%%%%%%%%%%%%%%%%%%%%%%%%%
\section{The Co-annihilation Scenario and Bound State Channels}\label{secII}
%%%%%%%%%%%%%%%%%%%%%%%%%%%%%%%%%%%%%%%%%%%%%%%

%%%%%%%%%%%%%%%%%%%%%%%%%%%%%%%%%%%%%%%%%%%%%%%
\subsection{The Boltzmann Equations}
%%%%%%%%%%%%%%%%%%%%%%%%%%%%%%%%%%%%%%%%%%%%%%%

We first give a general description of the co-annihilation scenario and include the bound state effects.
We call the DM particle $X$ and assume it is a SM singlet and conjugate to itself, and call its heavier partner particle $Y$ which is colored under the fundamental representation of $SU(3)_c$. In the cases we study, there is a tree level coupling between $X$, $Y$ and a third generation SM quark, denoted by $q$. Furthermore, the bound states (onia) formed by co-annihilation partners $Y$ and $\bar{Y}$ is denoted by $B$. During the DM freeze out, the following processes are relevant
\begin{itemize}
\item DM self-annihilation into SM (labelled by $XX$): $XX \leftrightarrow q\bar q$
\item DM-partner conversion (labelled by $X\leftrightarrow Y$): $Xq \leftrightarrow Yg$ and $Xg \leftrightarrow Y\bar q$, as well as decay and inverse decays of $Y\to X +{\rm SM \ particles}$ 
\item DM-partner co-annihilation (labelled by $XY$): $X Y \leftrightarrow q g$
\item Partner annihilation with Sommerfeld enhancement (labelled by $Y\bar Y$): $Y\bar Y \leftrightarrow gg$, $Y\bar Y \leftrightarrow q\bar q$, etc.
\item Bound state $B$ decay into DM particles (labelled by $B \leftrightarrow X$): $B \leftrightarrow XX$
\item Bound state $B$ decay into SM particles (labelled by $B \leftrightarrow {\rm SM}$): $B \leftrightarrow 2g$ or $3g$, $B \leftrightarrow q\bar q$, etc.
\item Bound state $B$ radiative capture and dissociation (labelled by $B \leftrightarrow Y$): $Y\bar Y \leftrightarrow Bg$
\end{itemize}
The coupled Boltzmann equations involving $X, Y, B$ are
\begin{eqnarray}
&&\hspace{-0.6cm}s H z \frac{d \mathcal{Y}_X}{d z} \!=\! - \gamma_{XX}\!\! \left[\! \left( \frac{\mathcal{Y}_X}{\mathcal{Y}_X^{eq}} \right)^2\! - 1 \right]\!
- \!2\gamma_{XY}\! \!\left[ \frac{\mathcal{Y}_X}{\mathcal{Y}_X^{eq}} \frac{\mathcal{Y}_Y}{\mathcal{Y}_Y^{eq}} - 1 \right]\!
- \!2\gamma_{X\leftrightarrow Y}\! \!\left[ \frac{\mathcal{Y}_X}{\mathcal{Y}_X^{eq}} - \frac{\mathcal{Y}_Y}{\mathcal{Y}_Y^{eq}} \right]\!
- \!2\gamma_{B\leftrightarrow X}\! \!\left[\! \left( \frac{\mathcal{Y}_X}{\mathcal{Y}_X^{eq}} \right)^2\! - \frac{\mathcal{Y}_{B}}{\mathcal{Y}_{B}^{eq}} \right]\!,\hspace{0.5cm} \label{YX} \\
&&\hspace{-0.6cm}s H z \frac{d \mathcal{Y}_Y}{d z} \!=\! - \gamma_{Y\bar Y}\! \!\left[\! \left( \frac{\mathcal{Y}_Y}{\mathcal{Y}_Y^{eq}} \right)^2 - 1 \right]\!
- \gamma_{XY}\! \!\left[ \frac{\mathcal{Y}_X}{\mathcal{Y}_X^{eq}} \frac{\mathcal{Y}_Y}{\mathcal{Y}_Y^{eq}} - 1 \right]\!
+ \gamma_{X\leftrightarrow Y}\! \!\left[ \frac{\mathcal{Y}_X}{\mathcal{Y}_X^{eq}} - \frac{\mathcal{Y}_{Y}}{\mathcal{Y}_{Y}^{eq}} \right]\!
- \gamma_{B\leftrightarrow Y}\! \!\left[\! \left( \frac{\mathcal{Y}_Y}{\mathcal{Y}_Y^{eq}} \right)^2 - \frac{\mathcal{Y}_{B}}{\mathcal{Y}_{B}^{eq}} \right]\!, \label{YY} \\
&&\hspace{-0.6cm}s H z \frac{d \mathcal{Y}_{B}}{d z} \!=\! 
- \gamma_{B\leftrightarrow X}\! \!\left[ \frac{\mathcal{Y}_{B}}{\mathcal{Y}_{B}^{eq}} - \left( \frac{\mathcal{Y}_X}{\mathcal{Y}_X^{eq}} \right)^2 \right]\!
- \gamma_{B\leftrightarrow Y}\! \!\left[ \frac{\mathcal{Y}_{B}}{\mathcal{Y}_{B}^{eq}} - \left( \frac{\mathcal{Y}_Y}{\mathcal{Y}_Y^{eq}} \right)^2 \right]\!
- \gamma_{B \leftrightarrow {\rm SM}}\! \!\left[ \frac{\mathcal{Y}_{B}}{\mathcal{Y}_{B}^{eq}} - 1 \right]\!. \hspace{3.5cm} \label{YB} 
\end{eqnarray}
In the above equations, $\mathcal{Y} \equiv n/s$ where $n$ is the number density of a species and $s$ is the entropy density in the universe. 
We assume Boltzmann distribution for calculating the thermal number densities.
The parameter $z \equiv m_X/T$ and $T$ is the temperature of the universe. The reaction rate $\gamma$ is defined as
\begin{eqnarray}\label{gamma}
\gamma \!&=&\! \int d\Phi_{initial} d\Phi_{final} (2\pi)^4 \delta^{4}(\Sigma p) |\mathcal{M}(initial \leftrightarrow final)|^2 e^{-(E_a+E_b)/T} \ .
\end{eqnarray}
This expression can be further simplified for the cases of decay and $2\to2$ annihilations. See appendix~\ref{app:gamma} for more detailed discussion on the form of $\gamma$ in decay and annihilation processes.
Clearly, $\gamma$ is identical for a process and its inverse process.

We assume CP conservation, $\mathcal{Y}_{Y}=\mathcal{Y}_{\bar Y}$ and $\gamma_{X\bar Y} = \gamma_{XY}$, $\gamma_{X\leftrightarrow \bar Y}=\gamma_{X\leftrightarrow Y}$, thus the Boltzmann equation for $\mathcal{Y}_{\bar Y}$ is the same as that for $\mathcal{Y}_Y$. This also explains the factor of $2$'s in Eq.~(\ref{YX}), which arises from $X$ co-annihilating with or converting into the anti-partner particle $\bar Y$. In contrast, there are no factor of 2 in front of $\gamma_{XX}$. It is canceled by a symmetry factor $1/2$  for identical $X$ particles in the initial state integral, as commented in~\cite{Griest:1990kh}.

The above coupled Boltzmann equations can be further simplified with the following considerations.
First, a useful observation is that during freeze out, the bound state formation and decay rates are all much larger than the Hubble expansion rate, {\it i.e.}, $\gamma_{B \leftrightarrow X, Y, g} \gg sHz$. This implies that it is a good approximation to set the value of $\mathcal{Y}_B$ equal to the quasi-static solution, which makes the right-hand side  of Eq.~(\ref{YB}) vanishing at every time,
\begin{eqnarray}\label{quasi}
\frac{\mathcal{Y}_{B}}{\mathcal{Y}_{B}^{eq}} \simeq \frac{
\gamma_{B\leftrightarrow X} \left( \frac{\mathcal{Y}_X}{\mathcal{Y}_X^{eq}} \right)^2
+\gamma_{B\leftrightarrow Y} \left( \frac{\mathcal{Y}_Y}{\mathcal{Y}_Y^{eq}} \right)^2
+\gamma_{B \leftrightarrow {\rm SM}}
}{\gamma_{B\leftrightarrow Y} + \gamma_{B\leftrightarrow X} + \gamma_{B \leftrightarrow {\rm SM}}} \ . 
\end{eqnarray}
Second, after the freeze out, the partner particles $Y, \bar Y$ will eventually decay into the DM $X$. Therefore, the final DM relic abundance is directly related to the quantity $\mathcal{Y}_{\rm dark} = \mathcal{Y}_X + \mathcal{Y}_Y + \mathcal{Y}_{\bar Y}$. The other useful observation is that, when the masses of $X$ and $Y$ are close enough,
the $X\leftrightarrow Y$ conversion rate is also much larger than the Hubble expansion rate during freeze out. This implies an approximate relation~\cite{Griest:1990kh},
\begin{eqnarray}\label{fast}
\frac{\mathcal{Y}_X}{\mathcal{Y}_X^{eq}} \simeq \frac{\mathcal{Y}_Y}{\mathcal{Y}_Y^{eq}} \ . 
\end{eqnarray}
Using Eqs.~(\ref{quasi}) and (\ref{fast}), we can add up the Boltzmann equations for $\mathcal{Y}_X, \mathcal{Y}_Y, \mathcal{Y}_{\bar Y}$ and simplify them into the following compact form,
\begin{eqnarray}\label{BEfinal}
s H z \frac{d \mathcal{Y}_{\rm dark}}{d z} = - \left[ \gamma_{XX} + 4 \gamma_{XY} + 2 \gamma_{Y\bar Y} 
+ 2\gamma_{\rm bound\ state}
\right] \left[ \left( \frac{\mathcal{Y}_{\rm dark}}{\mathcal{Y}_{\rm dark}^{eq}} \right)^2 - 1 \right] \ , 
\end{eqnarray}
where 
\begin{eqnarray}\label{gammaBS}
\gamma_{\rm bound\ state} = \sum_B \frac{(\gamma_{B\leftrightarrow X} + \gamma_{B\leftrightarrow Y})\gamma_{B \leftrightarrow {\rm SM}}}{\gamma_{B\leftrightarrow X} + \gamma_{B\leftrightarrow Y} + \gamma_{B \leftrightarrow {\rm SM}}} \ .
\end{eqnarray}

%%%%%%%%%%%%%%%%%%%%%%%%%%%%%%%%%%%%%%%%%%%%%%%
\subsection{Bound State Formation and Decays}\label{sectionIIb}
%%%%%%%%%%%%%%%%%%%%%%%%%%%%%%%%%%%%%%%%%%%%%%%

We first calculate the radiative bound state formation cross section for $Y \bar Y \to Bg$. 
Several remarks are in order.
First, because the $Y$ particle is a color triplet, the potential between $Y$ and $\bar Y$ is only attractive if they form a color singlet state. This means the bound state $B$ must be in a color singlet, and in turn, the initial state $Y\bar Y$ must be in a color octet state and they are repulsive to each other. We denote the difference between the initial and final state potential energy as 
\begin{eqnarray}\label{dv1}
\Delta V_1(r) = \frac{4\alpha_S^{(f)}}{3r} + \frac{\alpha_S^{(i)}}{6r} \ , 
\end{eqnarray}
where $\alpha_S^{(f)}$ is the strong coupling for the bound state, evaluated at the energy scale equal to the inverse Bohr radius $\mu = a_n^{-1}= 2\alpha_S m_Y/(3n)$ ($n$ is the principle quantum number)~\cite{Kats:2012ym}, and $\alpha_S^{(i)}$ is the strong coupling for the initial scattering state, evaluated at the energy scale equal to the momentum of the initial $Y$ particle in the center-of-mass frame. 
%This will impact the transition matrix element.
In addition, the color factor for radiating a gluon from the $Y$ or $\bar Y$ line is 4/3. This could be understood by sandwiching the transition matrix element $(T^A)_{ik}\delta_{jl}$ between the initial state $\sqrt{2} (T^{B})_{kl}$ and final state $\delta_{ij}/\sqrt{3}$, which gives $\delta^{AB}/\sqrt{6}$, where $i,j,k,l$ ($A,B$) are fundamental (adjoint) representation color indices. Squaring this yields 4/3. See appendix~\ref{app:color} for more details.
There is an additional factor 1/9 for averaging over the initial state colors. 
In appendix~\ref{app:dipole}, we give a brief discussion of how $\Delta V_1$ shows up in the above transition matrix element.

Second, thanks to the non-abelian nature of the $SU(3)_c$ gauge group, there is an additional diagram for the capture into bound states, where a real gluon is radiated from the Coulomb gluon exchanged between $Y$ and $\bar Y$. Effectively, it contributes as another potential term
\begin{eqnarray}\label{dv2}
\Delta V_2(r) = \frac{3\alpha_S}{2r} \ .
\end{eqnarray}
Here we evaluate $\alpha_s$ at the energy scale which is the average of the inverse Bohr radius and initial $Y$ particle momentum.
See appendix~\ref{app:nonabelian} for a detailed derivation of this contribution.
This piece of contribution was first noticed in \cite{Asadi:2016ybp} and \cite{Mitridate:2017izz}.
 
Under the dipole approximation, the general formation cross section of bound state formation in the $U(1)$ force case was derived in~\cite{An:2016gad}. Here, we generalize the result for the case of QCD bound state formation, with arbitrary quantum numbers $n, \ell$,
\begin{eqnarray}\label{colorkramers}
(\sigma v_{\rm rel})^{n, \ell}_{Y\bar Y \to Bg} = \frac{4\alpha_D}{81 \pi} \omega_{n} \left[ \ell \left|\int dr r^3 \left(\omega_{n} - \Delta V \right) R_{n\ell} R_{k \ell-1} \right|^2 
+(\ell+1) \left|\int dr r^3 \left(\omega_{n} - \Delta V \right) R_{n\ell} R_{k \ell+1} \right|^2 \right] \ ,
\end{eqnarray}
where $\omega_{n} = E_{n} + k^2/(2\mu)$ is the sum of the binding energy of the $n$'th bound state level and the kinetic energy of incoming state.
$R_{n\ell}$ and $R_{k \ell}$ are the radial part of the wavefunction of the bound state and initial scattering state, defined as
$\Psi_n({\bf r}) = \sum_{\ell m} R_{n\ell}(r) Y_{\ell m}(\hat r)$, $\Psi_k({\bf r}) = \sum_{\ell m} R_{k\ell}(r) Y_{\ell m}(\hat r) Y^*_{\ell m}(\hat k)$, respectively.
Here $\Delta V$ is the sum of the potential energies in Eqs.~(\ref{dv1}) and (\ref{dv2}),
\begin{eqnarray}
\Delta V(r) = \Delta V_1(r) + \Delta V_2(r) \ .
\end{eqnarray}
Because of the contribution from non-abelian gauge interaction, $\Delta V_2$, the above cross section does {\it not} agree with the analytic one given in~\cite{Ellis:2015vaa, Liew:2016hqo} for the case of ground state. In particular, an accidental cancellation occurs when evaluating the above matrix elements for the transition between a low energy scattering state to the ground state, as noticed in~\cite{Mitridate:2017izz}. In our calculations below, we take into account of the $1S$ and $2S$ bound states. The latter gives about 10-20\% correction to the former.

This above cross section is derived based on the Schr\"odinger equation with the spin-orbit interactions neglected at leading order, 
therefore, it is insensitive to the spin of the co-annihilating partner and identical between the scalar and fermionic partners. In the latter case, when Eq.~(\ref{colorkramers}) is used as the spin averaged cross section for calculating the reaction rate $\gamma_{Y\bar Y\to Bg}$,
three quarters of the formed bound states will be in spin triplet ($\Upsilon_Y$) and only one quarter in spin singlet ($\eta_Y$), {\it i.e.},
$\gamma_{Y\bar Y\to \Upsilon_Y g} = \frac{3}{4} \gamma_{Y\bar Y\to Bg}$, $\gamma_{Y\bar Y\to \eta_Y g} = \frac{1}{4} \gamma_{Y\bar Y\to Bg}$.

Next, we calculate the bound state decay rates via the $Y\bar Y$ annihilation inside it. We define the bound states made of both scalar and fermion colored partners in the appendix~\ref{app:bound}. For $S$-wave bound state decay, the general procedure is to first calculate the amplitude $\mathcal{A}_{ij}$ of the process $Y\bar Y \to$\,final state for $Y, \bar Y$ at rest with the external legs of $Y, \bar Y$ amputated ($i,j$ are their color indices). The bound state decay amplitude to the same final state is obtained by using the projection operators (see also~\cite{Petrelli:1997ge}),
\begin{eqnarray}
\mathcal{A}_{B\to {\rm final}} = 
\begin{cases}
\sqrt\frac{1}{3m_Y} \Psi(0) \delta_{ij} \mathcal{A}_{ij} \ , & \hspace{0.0cm} Y \ {\rm is \ scalar} \\
\sqrt\frac{1}{24m_Y^3} \Psi(0) \delta_{ij} {\rm Tr}\left[ \left(\frac{\not P}{2} + m_Y\right) \gamma_5 \left(\frac{\not P}{2} - m_Y\right) \mathcal{A}_{ij} \right] \ , & \hspace{0.0cm} Y \ {\rm is \ fermion, \ }B=\, \eta_Y \\
\sqrt\frac{1}{24m_Y^3} \Psi(0) \delta_{ij} {\rm Tr}\left[ \left(\frac{\not P}{2} + m_Y\right) \!\not\!\varepsilon(P) \left(\frac{\not P}{2} - m_Y\right) \mathcal{A}_{ij} \right] \ , & \hspace{0.0cm} Y \ {\rm is \ fermion, \ }B=\, \Upsilon_Y  \nonumber
\end{cases}
\end{eqnarray}
where $\Psi(0)$ is the bound state wavefucntion at the origin, $P^\mu$ is the four momentum of the bound state. In the above $\eta_Y$ denote the spin singlet (scalar) state while $\Upsilon_Y$ is the spin triplet (vector) state, and $\varepsilon$ is the polarization vector of the $\Upsilon_Y$ state.

With this approach, we calculate the $S$-wave bound state decay rates into gluons, which are (see also~\cite{Voloshin:2007dx}),
\begin{eqnarray}
&&\Gamma_{B\to gg} = 
\begin{cases}
\frac{4 \pi \alpha_S(m_Y)^2}{3m_Y^2} |\Psi(0)|^2 \ , & \hspace{0.0cm} Y \ {\rm is \ scalar} \\
\frac{8 \pi \alpha_S(m_Y)^2}{3 m_Y^2} |\Psi(0)|^2 \ , & \hspace{0.0cm} Y \ {\rm is \ fermion, \ }B=\, \eta_Y
\end{cases} \nonumber \\
&&\Gamma_{B\to ggg} = 
\frac{40 (\pi^2-9) \alpha_S(m_Y)^3}{81 m_Y^2} |\Psi(0)|^2 \ , \hspace{0.3cm} Y \ {\rm is \ fermion, \ }B=\, \Upsilon_Y
\end{eqnarray}
We neglect the electroweak decay channels of the bound states hereafter, which is subdominant and their effects are small.
We will present the bound state decay rates into DM particles $X$ in specific models in the next sections.

%%%%%%%%%%%%%%%%%%%%%%%%%%%%%%%%%%%%%%%%%%%%%%%
\section{Bino-Stop/Sbottom Co-annihilation}\label{secIII}
%%%%%%%%%%%%%%%%%%%%%%%%%%%%%%%%%%%%%%%%%%%%%%%

In this section, we consider more concrete models where the bino DM (i.e., an $SU(3)_c$ and $SU(2)_L$ singlet fermion) $\chi$ co-annihilates with a right-handed top or bottom scalar quark (which we will call stop or sbottom). This corresponds to the limit in MSSM where all other superpartners are much heavier and decoupled. Notice that this is similar to the assumption of the "simplified model" approach adopted in most direct searches for stop and sbottom at the LHC. Furthermore, the "co-annihilation" scenario where the bino and stop/sbottom are close in mass corresponds to the "compressed spectra" scenario \cite{LeCompte:2011cn}, where the limits from direct searches are significantly weakened.

%%%%%%%%%%%%%%%%%%%%%%%%%%%%%%%%%%%%%%%%%%%%%%%
\subsection{Bino-Stop}
%%%%%%%%%%%%%%%%%%%%%%%%%%%%%%%%%%%%%%%%%%%%%%%

The relevant interacting Lagrangian for the bino-stop case is
\begin{eqnarray}
\mathcal{L}_{\rm bino-stop} = \left( \frac{\sqrt2 g_1}{3} \bar t (1-\gamma_5) \chi \tilde t + {\rm h.c.} \right) + \frac{2m_t^2}{v^2} H^\dagger H \tilde t^* \tilde t \ ,
\end{eqnarray}
where $g_1$ is the hypercharge gauge coupling, $H$ is the Higgs doublet and $v=246\,$GeV. 
Applying the general discussion in the previous section to here, we have $X=\chi$ and $Y=\tilde t$.
With these interactions, we calculate the freeze out of bino DM in the presence of a slightly heavier stop, by including the bino self-annihilation process $\chi\chi \to t\bar t$, the bino-stop co-annihilation processes $\chi \tilde t \to t g, t h, t Z, t \gamma, b W$, the stop-anti-stop annihilation processes $\tilde t^* \tilde t \to gg, hh, ZZ, WW, 
\gamma\gamma, gh, gZ, g\gamma, hZ, g\gamma, Z\gamma$, as well as the stopponium bound state channels.
We calculate the Born-level cross sections of these processes using {\tt CalCHEP}~\cite{Belyaev:2012qa} which agree with the $S$-wave result presented in~\cite{Ibarra:2015nca}.
We also include the Sommerfeld enhancement/suppression factors discussed in~\cite{deSimone:2014pda, ElHedri:2016onc}.

%%%%%%%%%%%%%%%%%%%%%%%%%%%%%%%%%%%%%%%%%%%%%%%
\begin{figure}[t]
\centerline{\includegraphics[width=0.56\columnwidth]{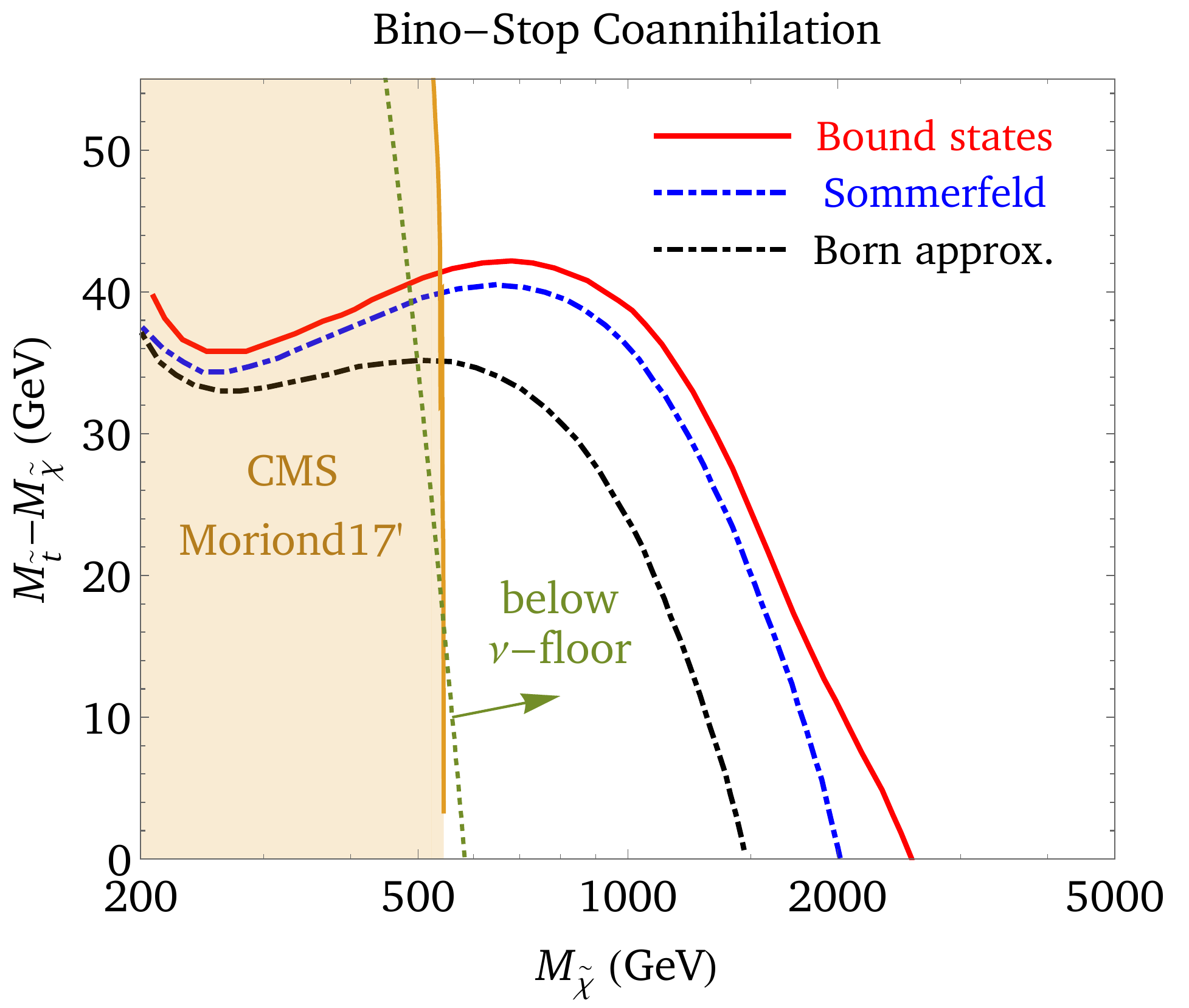}}
\caption{Parameter space where the bino dark matter could obtain correct relic abundance via co-annihilation with a right-handed top squark. The red solid, blue dot-dashed, black dot-dashed curves correspond to the freeze out calculation including both bound state and Sommerfeld effects, with only Sommerfeld effects, with neither, respectively.
%The orange and blue shaded regions are excluded by the current LHC results, while the future LHC reach is shown by the purple dashed curve. 
For direct detection, the region to the left of the green curve lies below the neutrino floor.}\label{binostop}
\end{figure}
%%%%%%%%%%%%%%%%%%%%%%%%%%%%%%%%%%%%%%%%%%%%%%%

For the bound state channels, the radiative stopponium formation cross section and its partial decay rate into two gluons have been calculated in the previous section. 
We consider the ground state in the calculation, whose quantum numbers are $J^{PC}=0^{++}$, and we call it $B=B_{\tilde t}$.
There are two additional decay channels, the bound state decay into two binos (via $t$-channel top quark exchange) and bound state decay into $t\bar t$ (via $t$-channel bino exchange). Because the bound state is a color singlet, the $s$-channel gluon exchange does not contribute to $B\to t\bar t$. Their partial decay width are
\begin{eqnarray}\label{B2bino}
\Gamma_{B_{\tilde t}\to \chi\chi} = \frac{128\pi \alpha_1^2 m_\chi^2 \left( m_{\tilde t}^2-m_{\chi}^2 \right)^{3/2}}{27m_{\tilde t}^3 \left(m_{\tilde t}^2-m_{\chi}^2+m_t^2 \right)^2}  |\Psi(0)|^2 \ , \nonumber \\
\Gamma_{B_{\tilde t}\to t\bar t} = \frac{64\pi \alpha_1^2 m_t^2 \left( m_{\tilde t}^2-m_{t}^2 \right)^{3/2}}{81m_{\tilde t}^3 \left(m_{\tilde t}^2-m_{t}^2+m_\chi^2 \right)^2}  |\Psi(0)|^2 \ .
\end{eqnarray}
where $\alpha_1=g_1^2/(4\pi)$.
For ground state $|\Psi(0)|^2=8 \alpha_S m_{\tilde t}^3/(27\pi)$ and the strong coupling $\alpha_S$ is evaluated at the energy scale equal to the inverse Bohr radius $\mu= 2\alpha_S m_Y/3$.
Both of the $B_{\tilde t}\to t\bar t$ and $B_{\tilde t}\to gg$ decay rates contribute to $\gamma_{B \leftrightarrow {\rm SM}}$ in the Boltzmann equation, and in Eq.~(\ref{gammaBS}). We neglected $B_{\tilde t}$ decays into $Z, \gamma, h$ channels because they are always subdominant to $B_{\tilde t}\to gg$.

%%%%%%%%%%%%%%%%%%%%%%%%%%%%%%%%%%%%%%%%%%%%%%%
\begin{figure}[t]
\centerline{\includegraphics[width=0.8\columnwidth]{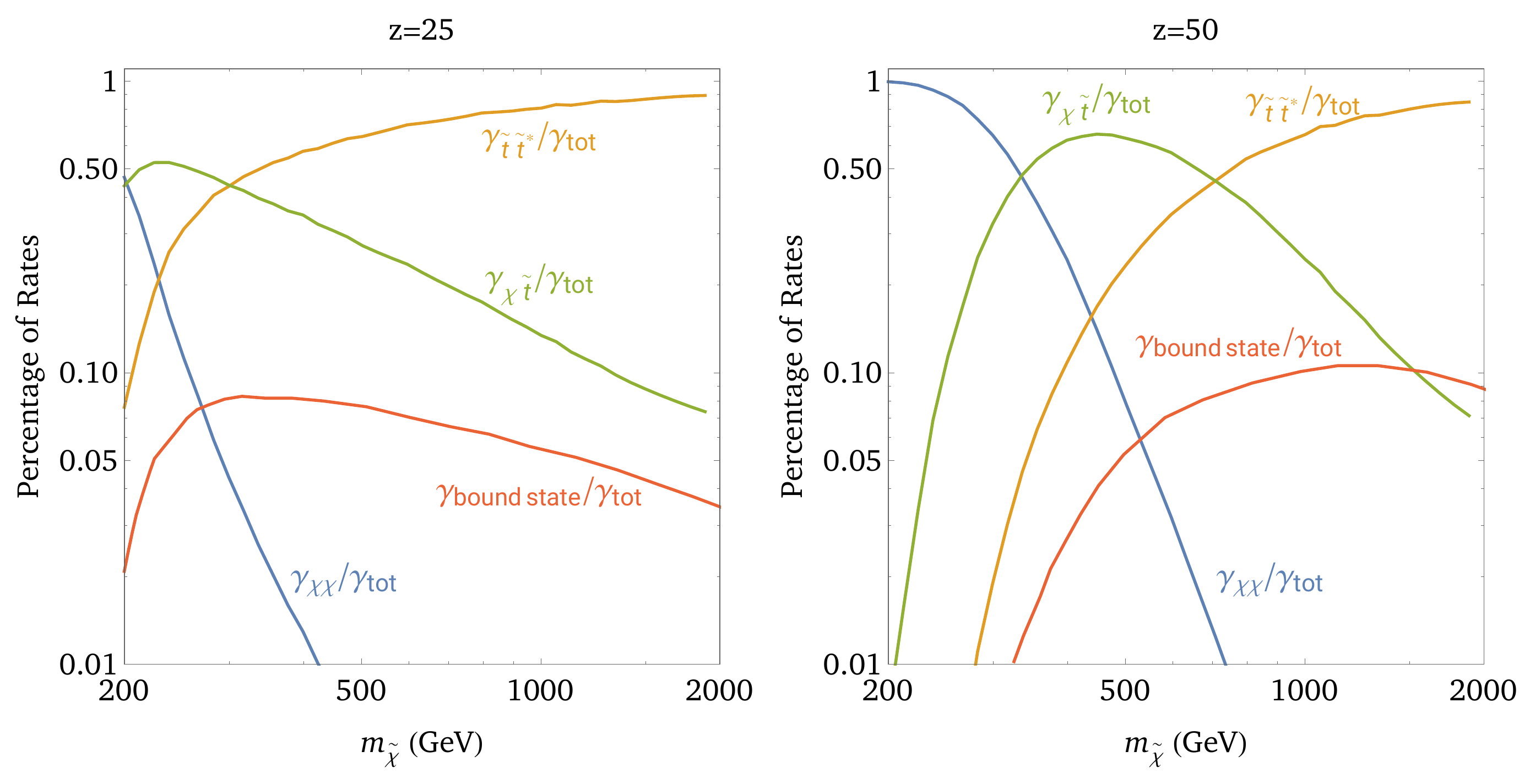}}
\caption{Importance of various reaction rates used in Eq.~(\ref{BEfinal}) at two different temperatures around the freeze out, where we have defined $\gamma_{\rm tot}=\gamma_{\chi\chi} + 4 \gamma_{\chi\tilde t} + 2 \gamma_{\tilde t\tilde t^*} + 2\gamma_{\rm bound\ state}$. We have fixed the stop-bino mass difference given by the red curve in Fig.~\ref{binostop}.
We find that, during freeze out, the contribution of bound state channels to the overall (co-)annihilation rate is always less than 10\%.
In particular, in the low bino mass region, the electroweak interactions involving the bino are very important, mainly because the bino number density is less Boltzmann suppressed than stop at low temperature.}\label{percentage}
\end{figure}
%%%%%%%%%%%%%%%%%%%%%%%%%%%%%%%%%%%%%%%%%%%%%%%

We solve the Boltzmann equation Eq.~(\ref{BEfinal}) with all the above processes included. The parameter space for bino-stop co-annihilation to yield the correct DM relic abundance is depicted in Fig.~\ref{binostop}.
The black dot-dashed curve corresponds to the calculation including Born-level stop-anti-stop annihilation and without bound state channels.
The blue dot-dashed curve corresponds to the calculation including stop-anti-stop annihilation with Sommerfeld effects but without bound state effects.
The red solid curves correspond to to the calculation with both Sommerfeld and bound state effects included.
Comparing the red and blue curves, we find that including the bound state effects could further increase the stop-bino mass difference by 5\,GeV for bino mass $m_\chi \lesssim1\,$TeV and as large as more than 10\,GeV for $m_\chi \sim2\,$TeV.

In our results, we find the over all stop-bino mass difference is  larger than that found in~\cite{Liew:2016hqo}.\footnote{Our results without the QCD bound-state effects, the dotted and the dashed curves in Fig.~\ref{binostop}, agree with those in Ref.~\cite{Ibarra:2015nca}.} On the other hand, 
the increase in the mass difference by including bound state effects (compare the red and blue curves) is less significant.
A crucial difference we notice is that~\cite{Liew:2016hqo} has only included the stop annihilation channels which are strong interactions but neglected all the electroweak processes, especially those involving the bino. For the bino-stop co-annihilation scenario, we find the electroweak interaction rates to be dominant over the QCD interaction rates during the freeze out for bino mass below $\sim$\,300-400\,GeV, and this is true for an even wider mass range at late times. This is mainly because the number density of stop (which is heavier) receives much more Boltzmann suppression than that of bino at temperatures smaller than the mass difference. 
Precision calculation of the co-annihilation mechanism requires taking into account of all the strong and electroweak interactions, as well as the non-perturbative Sommerfeld and bound state effects.
The importance of the electroweak interaction processes is shown in an example in Fig.~\ref{percentage}.
Once these processes are included, the bound state effects during freeze out is less than about 10\% throughout the bino mass range.

%%%%%%%%%%%%%%%%%%%%%%%%%%%%%%%%%%%%%%%%%%%%%%%
\begin{figure}[t]
\centerline{\includegraphics[width=0.56\columnwidth]{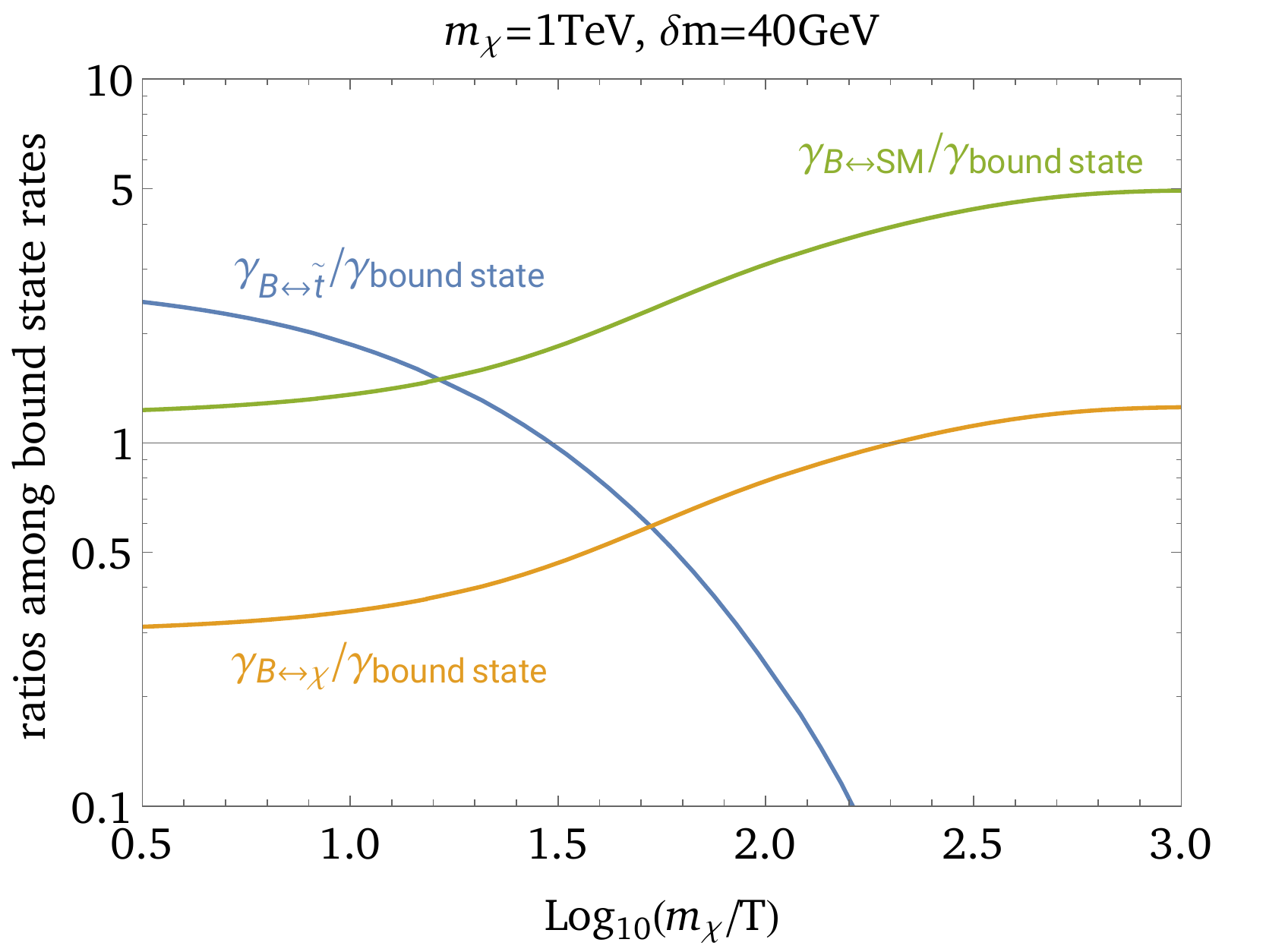}}
\caption{A comparison of detailed reaction rates of the stopponium bound state $B$ in unit of $\gamma_{\rm bound\ state}$, as a function of the temperature of the universe. We have chosen the bino mass to be 1\,TeV and fixed the stop-bino mass difference given by the red curve in Fig.~\ref{binostop}.
As discussed in the text, at higher temperature (left part of the plot), $\gamma_{\rm bound\ state} \simeq \gamma_{B \leftrightarrow {\rm SM}}$,
while a transition happens at low enough temperature so that $\gamma_{\rm bound\ state} \simeq \gamma_{B \leftrightarrow \chi}$.} 
\label{BSrate}
\end{figure}
%%%%%%%%%%%%%%%%%%%%%%%%%%%%%%%%%%%%%%%%%%%%%%%

Another bound state effect that has been neglected in the previous analysis is its decay into binos, Eq.~(\ref{B2bino}). This channel could have an impact on the behavior of the reaction rate $\gamma_{\rm bound\ state}$ at low temperatures of the universe.
When the temperature is large enough (before and around freeze out), we typically find the following hierarchy among the reaction rates in Eq.~(\ref{gammaBS}), $\gamma_{B\leftrightarrow \tilde t} \gg \gamma_{B \leftrightarrow {\rm SM}} \gg \gamma_{B\leftrightarrow \chi}$, and as a result, $\gamma_{\rm bound\ state} \simeq \gamma_{B \leftrightarrow {\rm SM}}$.
In contrast, at lower temperature (around and after freeze out), the hierarchy typically turns into
$\gamma_{B \leftrightarrow {\rm SM}} \gg \gamma_{B\leftrightarrow \chi}\gg \gamma_{B\leftrightarrow \tilde t}$, again because the number density of stop is more suppressed than bino. In this case, $\gamma_{\rm bound\ state} \simeq \gamma_{B \leftrightarrow \chi}$. There is a transition in the behavior of $\gamma_{\rm bound\ state}$ between these two regimes.
This feature is shown in an example in Fig.~\ref{BSrate}.

Because of the small mass difference in the stop-bino co-annihilation region, at the LHC the pair-produced stop has to decay into bino and through an off-shell top quark (and off-shell $W$-boson),
\begin{equation}
\tilde{t} \to t^* + \chi \to W^* + b + \chi \ ,
\end{equation}
The latest LHC constraint on such a decay mode is given in~\cite{moriond}, which excludes the bino mass up to $\sim 500\,$GeV for a compressed spectrum.
This is shown by the light orange shaded region in Fig.~\ref{binostop}.
Another potentially useful channel to search for the stop that is nearly degenerate with the bino is through stopponium production at colliders.  
A recent study shows the future running of high luminosity LHC could reach stop mass up to 350\,GeV via the stopponium\,$\to \gamma\gamma, Z\gamma$ decay channels~\cite{Batell:2015zla}.

The direct detection processes of bino DM in this case is generated at loop level. The box diagrams could give effective bino-gluon interactions and the triangle diagrams give bino-Higgs interactions. Both will contribute to the isospin-conserving and spin-independent scattering on the nucleon target.
We take both into account after correcting the sign of the Higgs contribution given in~\cite{Ibarra:2015nca}.
In Fig.~\ref{binostop}, in the region to the left (right) of the green dotted curve, the DM-nucleon scattering cross section is above (below) the neutrino floor.
It is worth noting that the current LHC stop search has almost ruled out all the parameter space above the neutrino floor.
In the future we must resort to the high energy/luminosity colliders for probing the remaining bino-stop co-annihilation parameter space.

%However, as noted in~\cite{Ibarra:2015nca}, a significant destructive interference between the box and the triangle diagrams occurs when the bino mass is close to the electroweak scale, due to the large top Yukawa coupling. (Such a cancellation does not happen for the bino-sbottom co-annihilation.)
%As a result, the scattering cross section in the region to the left of the green dotted curve in Fig.~\ref{binostop} is highly suppressed to be below the neutrino floor, which almost encloses all the bino-stop co-annihilation parameter space except for the sliver of region when the bino and the stop are almost degenerate at around 2.5 TeV. A message one could learn from these results is that if the future direct detection experiments find a positive signal of DM above the neutrino floor, the DM mass must be around 2.5\,TeV otherwise the above bino-stop co-annihilation scenario will be ruled out.

%%%%%%%%%%%%%%%%%%%%%%%%%%%%%%%%%%%%%%%%%%%%%%%
\subsection{Bino-Sbottonium}
%%%%%%%%%%%%%%%%%%%%%%%%%%%%%%%%%%%%%%%%%%%%%%%

%%%%%%%%%%%%%%%%%%%%%%%%%%%%%%%%%%%%%%%%%%%%%%%
\begin{figure}[t]
\centerline{\includegraphics[width=0.56\columnwidth]{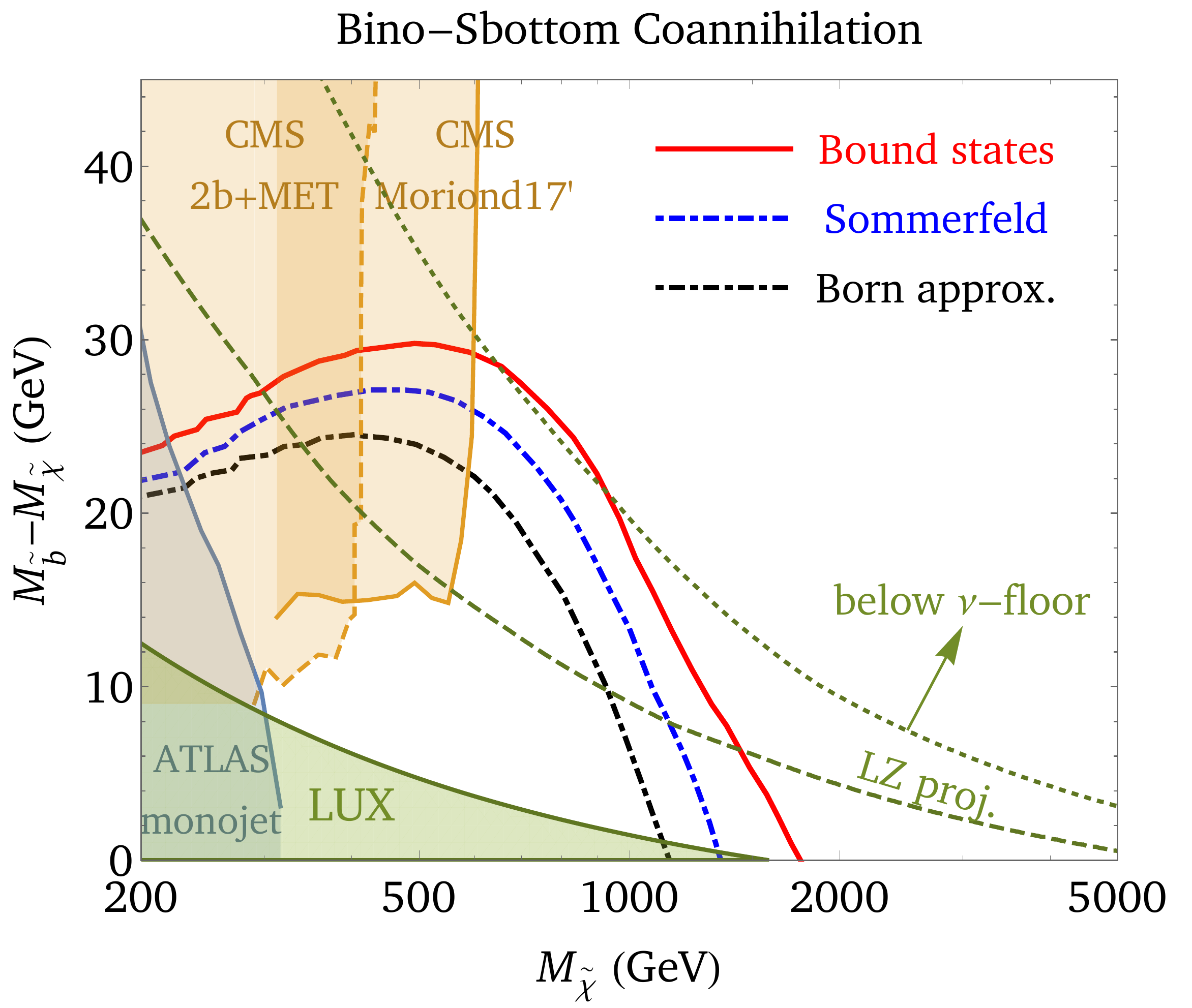}}
\caption{Parameter space where the bino dark matter could obtain correct relic abundance via co-annihilation with a right-handed bottom squark. The red solid, blue dot-dashed, black dot-dashed curves correspond to the freeze out calculation including both bound state and Sommerfeld effects, with only Sommerfeld effects, with neither, respectively.
The orange and blue shaded regions are excluded by the current LHC results.
The green shaded region is excluded by the latest dark matter direct detection result from LUX.
The direct detection cross sections corresponding to the future LZ experiment and the neutrino floor are given by the green dashed and dotted curves respectively.
An interesting finding here is that because of sbottomium bound state effects, part of the red curve that give the correct bino relic density turn out to hide below the neutrino floor.}\label{binosbottom}
\end{figure}
%%%%%%%%%%%%%%%%%%%%%%%%%%%%%%%%%%%%%%%%%%%%%%%

\begin{figure}[t]
\centerline{\includegraphics[width=0.8\columnwidth]{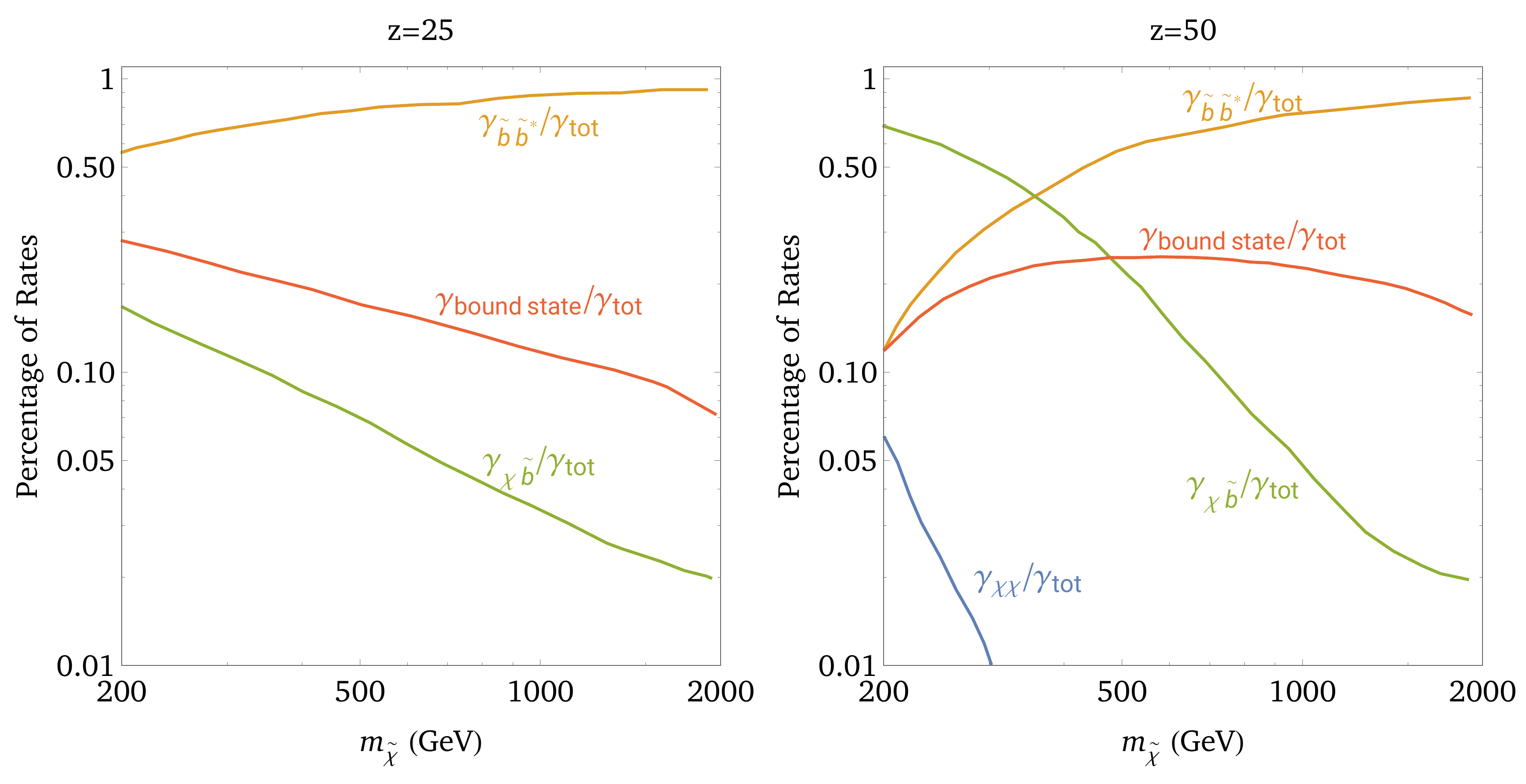}}
\caption{Same as Fig.~\ref{percentage} but for the bino-sbottom co-annihilation case.
The contribution of bound state channels to the overall (co-)annihilation rate during freeze out can be as large as 30\%, in the low bino mass region (the left part of the plot).}\label{sbottompercentage}
\end{figure}

Next, we discuss the case of bino-sbottom co-annihilation. The simplified Lagrangian in this case take the form
\begin{eqnarray}
\mathcal{L}_{\rm bino-sbottom} = \left( -\frac{\sqrt2 g_1}{6} \bar b (1-\gamma_5) \chi \tilde b + {\rm h.c.} \right) + \frac{2m_b^2}{v^2} H^\dagger H \tilde b^* \tilde b \ .
\end{eqnarray}
The bino-sbottom co-annihilation and freeze out is very similar to the bino-stop case, and early universe calculation results in both cases share many features discussed above, which will not repeated. The parameter space for bino-sbottom co-annihilation to yield the correct DM relic abundance is depicted in Fig.~\ref{binosbottom}.
We find the inclusion of the bound state channels has a stronger effect in enlarging the sbottom-bino mass difference than the stop-bino case. 
As shown in Fig.~\ref{sbottompercentage}, for low bino mass, the bound state channels could contribute as large as 30\% to the overall annihilation rate during freeze out.
This is mainly because the right-handed sbottom has smaller hypercharge than the stop, so the electroweak interactions involving the bino have smaller rates.

%%%%%%%%%%%%%%%%%%%%%%%%%%%%%%%%%%%%%%%%%%%%%%%
%%%%%%%%%%%%%%%%%%%%%%%%%%%%%%%%%%%%%%%%%%%%%%%

What we find the most interesting is the phenomenological implication of this result, especially in direct detection. In this case, because the bottom quark is light, the loop diagrams generating bino-gluon effective interaction (with sbottom and bottom running in the loop) is enhanced when the sbottom-bino mass difference is small~\cite{Drees:1993bu, Hisano:2010ct, Gondolo:2013wwa, Berlin:2015njh}.
The enhancement is the strongest when the mass difference is close to the bottom quark mass.
This is understood because the virtualness of the sbottom in the loop is controlled by this mass difference. The effective bino-gluon coupling is only suppressed by the mass square difference instead of the sbottom mass square.
In addition, there is no destructive interference like the stop case, because the effective bino-Higgs coupling is suppressed by the small bottom Yukawa coupling.
In Fig.~\ref{binosbottom}, the green shaded region with low bino mass and small mass difference is ruled out by the current LUX result~\cite{Akerib:2016vxi}.
The future LZ reach~\cite{Szydagis:2016few} as well as the neutrino floor are shown by the dashed and dotted green curves, respectively.
An important message one learns here is that the sbottomium bound state effects pushes part of the co-annihilation parameter space (the red solid curve) to lie below the neutrino floor. We will not be able to completely exclude the bino-sbottom co-annihilation scenario by doing the direct detection experiments only.  

It is interesting to note that the effects considered in this work give rise to a larger mass difference between the bino and its co-annihilating sbottom that would fit the observed relic density, which in turn results in a larger direct detection rate.  This is due to the fact that the one-loop  diagrams contributing to the gluon-bino interactions develops a pole after analytic continuation at \cite{Gondolo:2013wwa}  
\begin{equation}
\label{eq:pole}
m_{\tilde{b}} + m_b = m_\chi \ .
\end{equation}
The pole exists only when the gluon momentum vanishes, which is a very good approximation in direct detection. Since, by assumption,  $m_\chi <  m_{\tilde{b}}$, the pole is outside of the physical region of the scattering amplitudes. Nevertheless, since $m_b \ll m_\chi, m_{\tilde{b}}$, one could be sitting in the vicinity of the pole when the bino and sbottom are almost degenerate in mass and, therefore, receive an enhancement in the direct detection cross-section. Such an enhancement has been studied and demonstrated previously \cite{Gondolo:2013wwa,Berlin:2015njh}. These considerations makes it clear why the direct detection rate for bino-sbottom is reduced when the mass difference becomes larger, because one now moves further away from the pole in Eq.~(\ref{eq:pole}). 

In Fig.~\ref{binosbottom}, we also show the current LHC searches for sbottom decays,
\begin{equation}
\tilde{b} \to b + \chi \ ,
\end{equation}
which results in the $2b+$\,missing energy signature~\cite{CMSsbottom}, as well as the monojet channel~\cite{Aaboud:2016nwl, moriond} in which case the bottom quarks from sbottom decay are too soft so an initial state jet radiation is triggered on. As emphasized already, the direct searches at the LHC involve the same assumption as the co-annihilation scenario, that is the bino is close in mass to its co-annihilation partner, while all other superpartners are heavy and decoupled. We see that a bino with mass less than around 600 GeV is excluded by direct searches. With more data collected at the LHC, the direct searches will continue to have an impact on the allowed bino-sbottom co-annihilation region, and might eventually rule out the region where the direct detection rate lies below the neutrino floor.

%%%%%%%%%%%%%%%%%%%%%%%%%%%%%%%%%%%%%%%%%%%%%%%
\section{VDM-Top-Partner Co-annihilation}\label{secIV}
%%%%%%%%%%%%%%%%%%%%%%%%%%%%%%%%%%%%%%%%%%%%%%%

In this section, we consider another scenario with a vector dark matter (VDM) $V_\mu$ co-annihilating with a heavier fermionic top quark partner $T$. 
This could be motivated by extra dimensional models where the DM stability is related to a KK parity \cite{Hisano:2010yh}.~\footnote{Strictly speaking, the spectrum required for the co-annihilation studied in this section does not appear in the usual minimal UED models \cite{Servant:2002aq}. However, one could potentially introduce brane-localized kinetic terms to generate the desired mass spectrum \cite{Carena:2002me,Agashe:2007jb}.}
We take a simplified model approach and assume the following renormalizable interacting Lagrangian is
\begin{eqnarray}\label{VDM-TP}
\mathcal{L}_{\rm VDM-TP} = \left( \bar T  \gamma^\mu \left( \lambda_{VTt} + \lambda'_{VTt} \gamma_5 \right) t V_\mu + {\rm h.c.} \rule{0mm}{5mm}\right) + \lambda_{VVHH} V_\mu V^\mu H^\dagger H \ .
\end{eqnarray}
In the following discussions we will assume $\lambda_{VTt}$ to be real and set $\lambda'_{VTt}=0$ for simplicity.\footnote{Modifying this assumption will not alter qualitatively the results shown below.} Furthermore, we leave both $ \lambda_{VTt} $ and $\lambda_{VVHH}$ as free parameters in the model.

In the co-annihilation calculation, we have included $VV\to t\bar t, hh$ for DM self annihilation, $VT\to gt$ for DM-partner coannihilation, and $T\bar T\to gg, q\bar q$ taking into account the Sommerfeld effects from QCD interaction. The $t$-channel exchange of $V$ particle can also mediate the same-sign annihilation of $TT\to tt$. We have also included the bound state channels. Here because the top partner $T$ is a fermion, there are two ground states made of $T\bar T$,~\footnote{It is also possible for $TT$ to bind into same sign bound state if they form a color sextet~\cite{Giacchino:2015hvk}. Such a bound state could annihilate decay into two same-sign top quarks. Numerically, we find the contribution from such a bound state to co-annihilation is negligible, because it has a weaker Coulomb potential ($V=-\alpha_S/(3r)$) and it has to decay via the weak coupling $\lambda_{VTt}$.}
spin zero state $\eta_T$ and spin one state $\Upsilon_T$. Their formation cross sections and partial decay rates into gluons have been discussed in Section~\ref{sectionIIb}. In addition, the bound state could also decay into two vector DM particles (via $t$-channel top quark exchange), and into $t\bar t$ (via $t$-channel $V$ exchange). These decay rates are
\begin{eqnarray}
\Gamma_{\eta_T\to VV} \!\!\!&=&\!\!\! \frac{3 \lambda_{VTt}^4 (m_T^2-m_V^2)^{3/2}}{\pi m_T (m_T^2-m_V^2+m_t^2)^2} |\Psi(0)|^2  \ , \nonumber \\
\Gamma_{\eta_T\to t\bar t} \!\!\!&=&\!\!\! \frac{\lambda_{VTt}^4 \left[-2 m_V^2(2m_T-m_t) 
+ (m_T+m_t)(m_T-m_t)^2 \right]^2\sqrt{m_T^2 - m_t^2}}{8\pi m_T m_V^4 (m_T^2-m_t^2+m_V^2)^2} |\Psi(0)|^2  \ .
\end{eqnarray}
Next, for $\Upsilon_T$, in the absence of the $\lambda'_{VTt}$ coupling, the decay $\Upsilon_T\to VV$ is forbidden by $C$-parity under the assumption $\lambda'_{VTt}=0$.~\footnote{The $\Upsilon_T\to VV$ decay channel can be turned on with $\lambda_{VTt}$ and $\lambda'_{VTt}$ both nonzero, which break $C$-parity. 
In practice, we find that it makes very little change to the overall bound state effect, because most of the $\Upsilon_T$ that are formed are dissociated at much larger rate than decay.}
The decay rate of $\Upsilon_T\to t\bar t$ is,
\begin{eqnarray}
\Gamma_{\Upsilon_T\to t\bar t} \!\!\!&=&\!\!\! \frac{\lambda_{VTt}^4\sqrt{m_T^2 - m_t^2}}{6\pi m_T (m_T^2-m_t^2+m_V^2)^2} |\Psi(0)|^2 \nonumber \\
\!\!\!&&\!\!\! \times\left[ (2m_T^2+m_t^2) - \frac{(m_T+m_t) (2 m_T+ m_t) (m_T - m_t)^2}{m_V^2}
+\frac{3 (m_T + m_t)^2 (m_T - m_t)^4}{4m_V^4}
\right] \ .
\end{eqnarray}

\begin{figure}[h!]
\centerline{
\includegraphics[width=0.42\columnwidth]{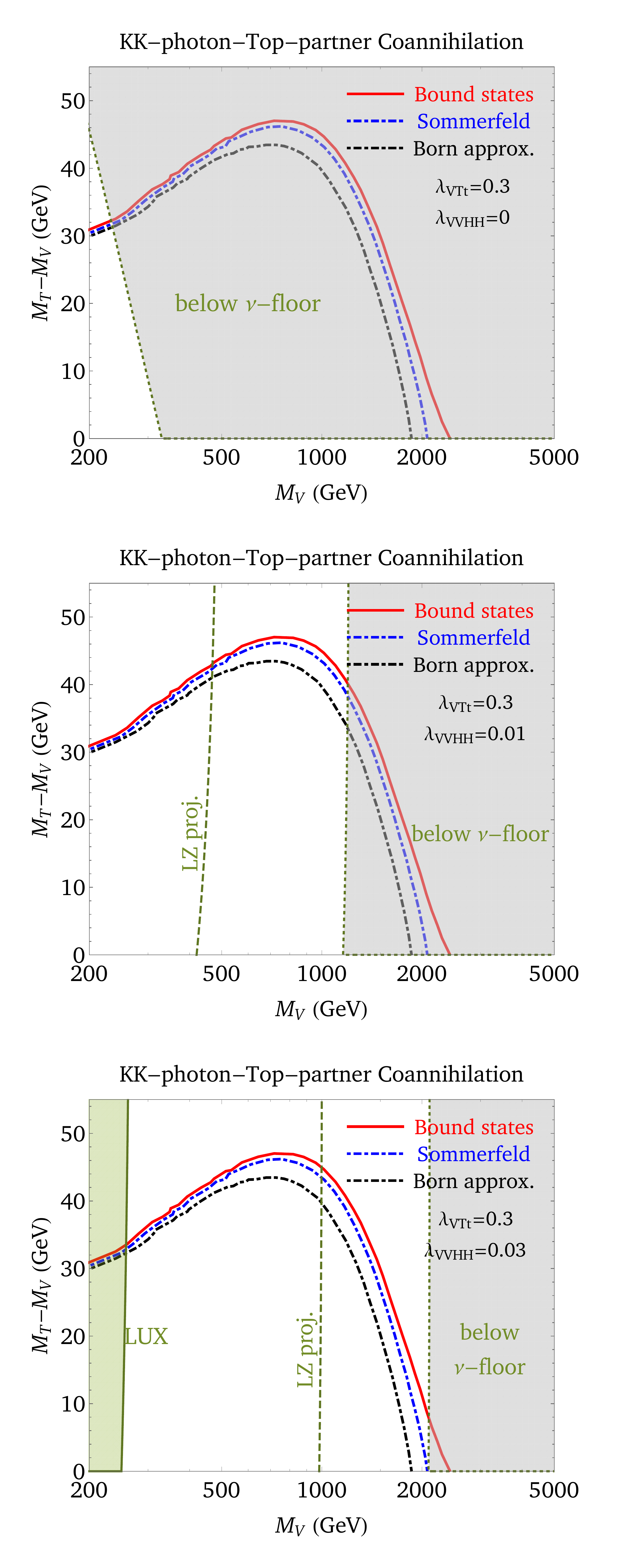}\hspace{0.5cm}
\includegraphics[width=0.42\columnwidth]{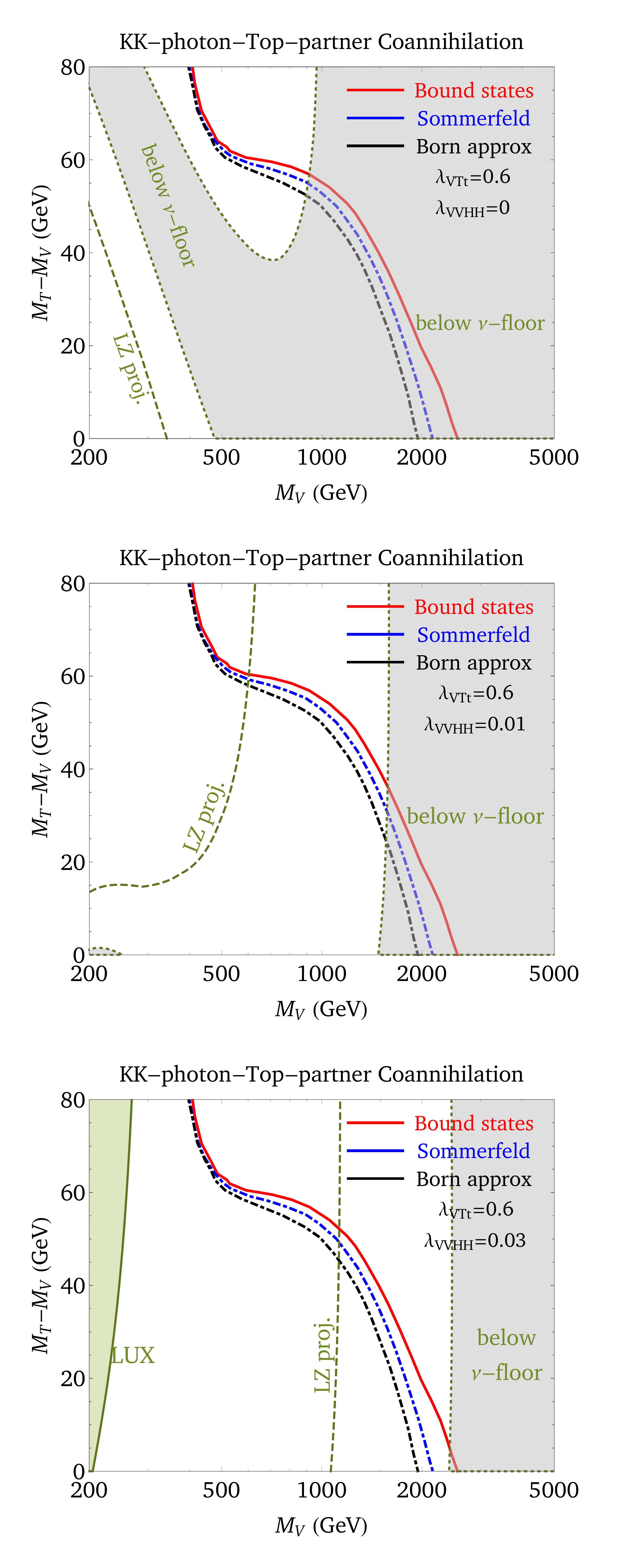}
}
\caption{Parameter space where a vector boson dark matter could obtain correct relic abundance via co-annihilation with a fermionic top quark partner, for various value of
$\lambda_{VTt}$ and $\lambda_{VVHH}$ parameters defined in Eq.~(\ref{VDM-TP}). The red solid, blue dot-dashed, black dot-dashed curves correspond to the freeze out calculation including both bound state and Sommerfeld effects, with only Sommerfeld effects, with neither, respectively.
The green shaded region is excluded by the latest dark matter direct detection result from LUX.
The dark matter direct detection cross section corresponding to the future LZ experiment are given by the green dashed curves.
The gray shaded regions are the parameter space with direct detection cross section below the neutrino floor.
}\label{KK}
\end{figure}

Our results are shown in Fig.~\ref{KK}, for  $\lambda_{VTt}=\{0.3, 0.6\}$ and $\lambda_{VVHH}=\{0, 0.01,0.03\}$, respectively.
For  $\lambda_{VTt}=-0.6$, we find that for    $m_V \lesssim 400\,$GeV, the VDM self annihilation cross section alone is large enough to explain the relic abundance, which in turn
allows the mass difference between top-partner and VDM to be very large.
We find the bound state effect on the top-partner-VDM mass difference is smaller than the squark-bino case. 
One important reason is that 3/4 of the bound states that are formed in this case are in the $\Upsilon_T$ state, which unlike $\eta_T$, always has to decay into three gluons because of the Landau-Yang theorem (note $\Upsilon_T$ is a color-singlet bound state), which is too slow. 
As a result, most of the $\Upsilon_T$ that are formed during the freeze out temperature are quickly dissociated before it has time to decay into gluons.
More quantitatively, Eq.~(\ref{gammaBS}) tells that the contribution to the bound state rate from $\Upsilon_T$ is much smaller than $\eta_T$.

%%%%%%%%%%%%%%%%%%%%%%%%%%%%%%%%%%%%%%%%%%%%%%%
%%%%%%%%%%%%%%%%%%%%%%%%%%%%%%%%%%%%%%%%%%%%%%%

Next we discuss the interplay of our result with the DM direct detection searches, where the current LUX exclusion, the future LZ reach and the neutrino floor are shown by
the green shaded region, green dashed and dotted curves. 
The spin-independent VDM-nucleon scattering could happen via the $V$-gluon effective coupling which is generated at loop level~\cite{Hisano:2010yh}, and the $VVHH$ interaction.~\footnote{In addition to the tree level $\lambda_{VVHH} $ term in Eq.~(\ref{VDM-TP}),
there is also a loop level contribution to the $VVHH$ coupling which involves the top quark and the top partner. We find the loop contribution is divergent which means that it makes the tree level coupling run with energy scales. In our calculation, the value of $\lambda_{VVHH}$ we use corresponds to the one after this renormalzation.}
We vary the value of $\lambda_{VVHH}$ within the range from 0 to 0.03. 
The upper limit is motivated by the coupling between Higgs and the first KK hypercharge gauge boson in the minimal UED, which is $g_1^2/4\simeq0.03$.
We find that for $\lambda_{VVHH}=0$, there is a cancellation between different box diagrams contributing to the  DM-gluon effective operator (as noted in~\cite{Chen:2016rae}), which results in a big part of the parameter space (shown by the gray shaded region in Fig.~\ref{KK}) dipping below the neutrino floor of direct detection experiments.
Such a cancellation is much less visible when we turn on a large enough value of $\lambda_{VVHH}$.
In contrast, varying the $\lambda_{VVHH}$ coupling in the above range has little impact on the co-annihilation curves. 

\renewcommand{\thefootnote}{\roman{footnote}}

On the other hand, at the LHC the only decay channel for the fermionic top partner in our simplified model is
\begin{equation}
T \to t + V \ ,
\end{equation}
where the dark matter particle $V$ manifests itself as missing transverse energy (MET). However, as can be seen from Fig.~\ref{KK}, the required mass difference between $V$ and $T$ to explain the observed relic density is quite small, less than the mass of the $W$ boson, which implies $T$ can only decay through a highly off-shell top into $b$-quark and an off-shell $W$ boson. The resulting signature at the LHC for the pair production~\footnote{The top partner needs to be pair-produced because they are odd under KK-parity.} of $T$ is therefore soft jets, soft leptons and a large amount of MET. We are not aware of any public results for searches for fermionic top partners in this particular channel. Needless to say, it would interesting to pursue this channel in future searches.

%%%%%%%%%%%%%%%%%%%%%%%%%%%%%%%%%%%%%%%%%%%%%%%
\section{Conclusion}\label{secV}
%%%%%%%%%%%%%%%%%%%%%%%%%%%%%%%%%%%%%%%%%%%%%%%

To summarize, in this work, we perform a precision calculation of dark matter relic abundance through the co-annihilation mechanism in the context of a few simple models, where the dark matter particle (assumed to be a SM gauge singlet) is accompanied by a slightly heavier and colored top/bottom partner particle. We consider the cases where the quantum numbers of the two dark particles allow a renormalizable interaction with a standard model quark. Given this interaction, the dark matter particles could annihilate during freeze out into standard model particles by: 1) being converted into the partner and then the partners annihilate away through strong interactions, and 2) direct annihilating into quarks through a $t$- and $u$-channel partner exchange or co-annihilate with a partner particle. In the partner-anti-partner annihilation channels belonging to 1), we have include the non-perturbative Sommerfeld as well as the QCD bound state effects in our calculation. In contrast, the annihilation channels belong to 2) are sometimes neglected in previous studies. We point out the importance of these channels --- they must be taken into account for a precision calculation of freeze out. Based on these considerations, we derive the cosmologically favored mass difference between the dark matter and its partner.

We confront the derived parameter space for dark matter relic abundance with the present and future experimental searches, which reveals several interesting interplay. For the case of bino-right-handed-stop co-annihilation, most of the parameter space is above the neutrino floor in direct detection has already been excluded by the current LHC stop search. Therefore, we have to rely on the future LHC and high energy colliders instead of direct detection experiments to further probe such a dark matter candidate. On the other hand, for the case of bino-right-handed-bottom co-annihilation, we find the direct detection experiments can serve as powerful probes. However, including the bound state effects in co-annihilation pushes part of the parameter space (with dark matter mass between 600\,GeV and 1\,TeV) to be below the neutrino floor. In this region, the collider searches could play a complementary role.

Our analysis can be implemented to various models with nearby colored partners of the dark matter in a straightforward manner. 
For example, we include a case study of the vector dark matter with a fermionic top quark partner.

%%%%%%%%%%%%%%%%%%%%%%%%%%%%%%%%%%%%%%%%%%%%%%%
\section*{Acknowledgement}
%%%%%%%%%%%%%%%%%%%%%%%%%%%%%%%%%%%%%%%%%%%%%%%

This work is supported by the DOE grant DE-SC0010143 at Northwestern, DOE grant DE-FG-02-12ER41811 at UIC and DOE grant DE-AC02-06CH11357 at ANL. We acknowledge useful conversations with G. Bodwin and C. Wagner. We also thank the authors of Refs.~\cite{Mitridate:2017izz,ElHedri:2017nny} for communications which led to improvements of our work.

%%%%%%%%%%%%%%%%%%%%%%%%%%%%%%%%%%%%%%%%%%%%%%%
\appendix
%%%%%%%%%%%%%%%%%%%%%%%%%%%%%%%%%%%%%%%%%%%%%%%

%%%%%%%%%%%%%%%%%%%%%%%%%%%%%%%%%%%%%%%%%%%%%%%
\section{Thermal Reaction Rates in Boltzmann Equations}\label{app:gamma}
%%%%%%%%%%%%%%%%%%%%%%%%%%%%%%%%%%%%%%%%%%%%%%%

In this appendix, we give the expressions for the thermal averaged reaction rate $\gamma$ defined in Eq.~(\ref{gamma}), for $1\to2$ decay and $2\to2$ annihilation processes.
It is worth noting that $\gamma$ is the same for a process and its inverse process. 

For a decay process $a\to c+d$, we have
\begin{eqnarray}\label{gamma1to2}
\gamma_{a\leftrightarrow cd} = \frac{T m_a^2}{2\pi^2} K_1\!\left( \frac{m_a}{T} \right) g_a \Gamma_{a\to cd} \ , 
\end{eqnarray}
where $K_1$ is the Bessel function of the first kind and $\Gamma_{a\to cd}$ is the partial decay rate in the rest frame of $a$.

For an annihilation process $a+b\to c+d$, 
\begin{eqnarray}\label{gamma2to2}
\gamma_{ab\leftrightarrow cd} = \frac{T}{8\pi^4} \int_{s_{\rm th}}^\infty ds \sqrt{s} K_1\!\left( \frac{\sqrt{s}}{T} \right) |{\bf p}_{a}^{\rm cm}|^2 g_a g_b\, \sigma_{a+b\to c+d} \ , 
\end{eqnarray}
where $s_{\rm th} = {\rm Max}[(m_a+m_b)^2, (m_c+m_d)^2]$ and $|{\bf p}_{a}^{\rm cm}|%=|{\bf p}_{b}^{\rm cm}| 
= \sqrt{[s-(m_a+m_b)^2][s-(m_a-m_b)^2]}/(2\sqrt{s})$.
When the initial state particle masses are equal to each other, {\it i.e.}, $m_a=m_b$, the above formula can be simplified to
\begin{eqnarray}\label{gamma2to2eqmass}
\gamma_{ab\leftrightarrow cd} = \frac{T}{64\pi^4} \int_{s_{\rm th}}^{\infty} ds s \sqrt{s-4m_a^2} K_1 \left(\frac{\sqrt s}{T}\right) \times g_a g_b\,  (\sigma v_{\rm rel})_{ab\to cd} \ , 
\end{eqnarray}
where $v_{\rm rel}$ is the relative velocity between $a$ and $b$.

%%%%%%%%%%%%%%%%%%%%%%%%%%%%%%%%%%%%%%%%%%%%%%%
\section{Color Factors}\label{app:color}
%%%%%%%%%%%%%%%%%%%%%%%%%%%%%%%%%%%%%%%%%%%%%%%

In this appendix, we discuss several processes of colored partner $Y$ and $\bar Y$ interaction, and derive the corresponding color factors in the corresponding amplitudes.

We first define the states. For a $Y_i\bar Y_j$ system where $i,j$ are the color indices, the color singlet state is obtained by contraction with $|1\rangle_{ij} = \delta_{ij}/\sqrt{3}$, while the color octet state is obtained by contraction with $|8_A\rangle_{ij} = \sqrt{2} (T^A)_{ij}$. The normalization of these factors are obtained by requiring their square to yield the color multiplicity.

The first process we consider is a photon emission in $Y\bar Y$ scattering. Clearly, their color indices do not change in such a process, so the color part of the transition operator is simply $\delta_{ik}\delta_{jl}$. The color factor for color singlet to singlet transition is then $_{ij}\langle 1|\delta_{ik}\delta_{jl}|1\rangle_{kl} = 1$.
The color factor for color octet to octet transition is then $_{ij}\langle 8_A|\delta_{ik}\delta_{jl}|8_B\rangle_{kl} = \delta_{AB}$. This is a trivial example.

Second, we consider a gluon emission in $Y\bar Y$ scattering. In this case, because the gluon is a color octet, one of the initial and final states must be a color octet and the other is color singlet. The transition operator for gluon radiated from the $Y$ particle is $(T^B)_{ik} \delta_{jl}$. And the color factor for octet to singlet transition is
$_{ij}\langle 1|(T^B)_{ik}\delta_{jl}|8_A\rangle_{kl} = \delta_{AB}/\sqrt6$. Squaring the amplitude and sum over $A, B$ yields the color factor 4/3, which is used for deriving Eq.~(\ref{colorkramers}).

Third, we consider the Coulomb potential between $Y$ and $\bar Y$ generated by a gluon exchange. In this case the operator is $(T^A)_{ik} (T^A)^*_{jl} = (T^A)_{ik} (T^A)_{lj}=\frac{1}{2} \delta_{ij}\delta_{kl} - \frac{1}{6} \delta_{ik}\delta_{jl}$. For the Coulomb potential between $Y\bar Y$ in a color singlet, we have the color factor, $_{ij}\langle 1|(T^A)_{ik} (T^A)_{lj}|1\rangle_{kl} = 4/3$.
For the Coulomb potential between $Y\bar Y$ in a color octet, the color factor is, $_{ij}\langle 8_B|(T^A)_{ik} (T^A)_{lj}|8_C\rangle_{kl} = -\delta_{BC}/6$.
This explains why $Y\bar Y$ in a color singlet (octet) state are attractive (repulsive) to each other, as well as the normalization of the gluon-exchange Coulomb potentials.

%%%%%%%%%%%%%%%%%%%%%%%%%%%%%%%%%%%%%%%%%%%%%%%
\section{Capture into Bound States: Gluon Radiated from $Y, \bar Y$}\label{app:dipole}
%%%%%%%%%%%%%%%%%%%%%%%%%%%%%%%%%%%%%%%%%%%%%%%

%%%%%%%%%%%%%%%%%%%%%%%%%%%%%%%%%%%%%%%%%%%%%%%
\begin{figure}[h]
\centerline{
\includegraphics[width=0.8\columnwidth]{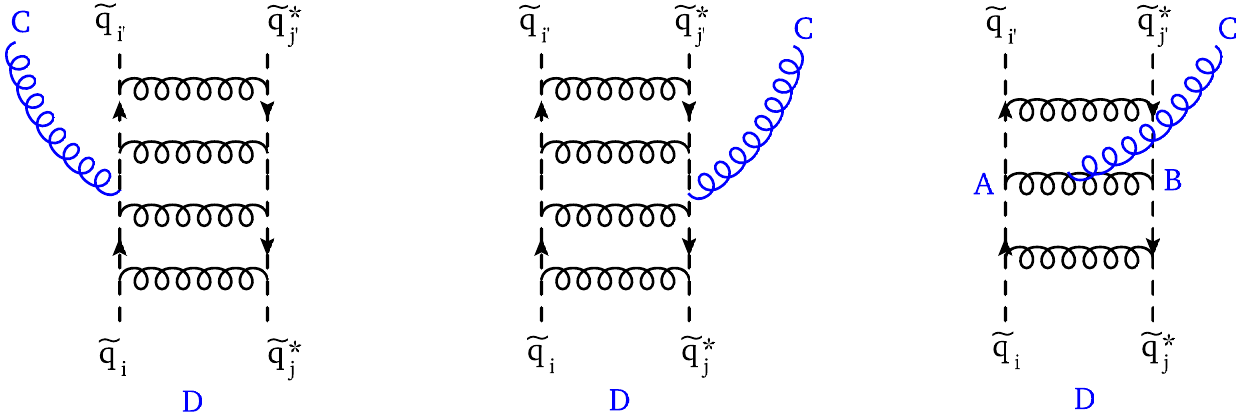}}
\caption{Feynman diagrams for capture into bound states by radiating a gluon. $A,B,C,D$ are the color indices in the adjoint representation.}\label{radiation}
\end{figure}
%%%%%%%%%%%%%%%%%%%%%%%%%%%%%%%%%%%%%%%%%%%%%%%

In the process $Y_i \bar Y_j \to Y_{i'} \bar Y_{j'}$ ($i, j, i', j'$ are the color indices), the effective Hamiltonian for an ultra-soft gluon (with color index $C$) radiated from $Y$ or $\bar Y$ is
\begin{eqnarray}
H_{int}^{(1)} = \frac{g_S  T^C_{i'i} \delta_{j'j}}{m_Y} \vec{k} \cdot \vec{G}^C(\vec{r}/2) + \frac{g_S  \delta_{i'i} T^C_{j'j}}{m_Y} \vec{k} \cdot \vec{G}^C(-\vec{r}/2)
\end{eqnarray}
where $\vec{k}$ is the relative momentum operator between $Y$ and $\bar Y$, and $G$ is the gluon field. The matrix element of $H_{int}^{(1)}$ between the final bound state (color singlet) and the initial scattering state (color octet, carrying adjoint color index $D$) is then (see the first two diagrams in Fig~\ref{radiation}),
\begin{eqnarray}
V_{fi}^{(1)} = \frac{\delta^{CD}}{\sqrt6}\frac{g_S}{\mu} \langle \Psi_f | \vec{k} | \Psi_i \rangle \cdot \vec{\varepsilon}^{\,C}_{\lambda} \ ,
\end{eqnarray}
where $\mu=m_Y/2$, and the color factor in the front is obtained based on the discussions in appendix~\ref{app:color}. 
$\varepsilon^A_{\lambda}$ is the polarization vector of the radiated gluon.
The wavefunctions $\Psi_f$ and $\Psi_i$ are the eigenstates of the Hamiltonians $\hat H_1 = \vec{k}^2/(2\mu) - 4 \alpha_s^{(f)}/(3r)$, and $\hat H_8 = \vec{k}^2/(2\mu) + \alpha_s^{(i)}/(6r)$, respectively, with
$\hat H_1 |\Psi_f\rangle = E_f |\Psi_f\rangle$ and $\hat H_8 |\Psi_i\rangle = E_i |\Psi_i\rangle$. 
Here $\alpha_S^{(f)}$ is the strong coupling for the bound state, evaluated at the energy scale equal to the inverse Bohr radius, and $\alpha_S^{(i)}$ is the strong coupling for the initial scattering state, evaluated at the energy scale equal to the momentum of the initial $Y$ particle in the center-of-mass frame. 
We define $\Delta V_1 = \hat H_8 - \hat H_1$, which yields Eq.~(\ref{dv1}).

Using $\vec{k} = - i \mu [\vec{r}, H_8] = - i \mu \left(\vec{r} H_8 - H_1 \vec{r} - \Delta V_1(r) \vec{r} \rule{0mm}{3.5mm}\right)$, we get
\begin{eqnarray}\label{vfi1}
V_{fi}^{(1)} = \frac{\delta^{CD}}{\sqrt6} (-i g_S) \langle \Psi_f | \left( E_i - E_f - \Delta V_1(r) \right) \vec{r} | \Psi_i \rangle \cdot \vec{\varepsilon}^{\, C}_{\lambda} \ .
\end{eqnarray}

%%%%%%%%%%%%%%%%%%%%%%%%%%%%%%%%%%%%%%%%%%%%%%%
\section{Capture into Bound States via Non-abelian Gauge Interaction}\label{app:nonabelian}
%%%%%%%%%%%%%%%%%%%%%%%%%%%%%%%%%%%%%%%%%%%%%%%

Because the gluons have their own self-interactions, it is also possible to radiate the gluon from a gluon propagator exchanged between $Y,\ \bar Y$ particles, as shown by the third diagram of Fig~\ref{radiation}. In non-relativistic limit, at leading order, the gluon being exchanged is a soft Coulomb gluon. In this case, the interaction between a radiative gluon and two Coulomb gluons can be written as
\begin{eqnarray}
\mathcal{L}_{3G} = g_S f^{ABC} (\vec{\nabla G_0^A}) G_0^B \cdot \vec{G}^C \ ,
\end{eqnarray}
where $f^{ABC}$ is the $SU(3)$ group structure constant. The corresponding covariant derivative on the fundamental $Y$ field is, $D_\mu Y_i = \partial_\mu Y_i - i g_S G^A_\mu T^A_{ij} Y_j$. With these, we calculate the amplitude for the third diagram of Fig~\ref{radiation},
\begin{eqnarray}
\mathcal{M} = 2 (i g_S T^A_{i'i}) \left( \frac{-i}{q^2} \right) (i g_S f^{ABC}) (i \vec{q} \cdot \vec{\varepsilon}^{\,C}) \left( \frac{-i}{q^2} \right) (-i g_S T^B_{j'j})
= 2 g_S^3 T^A_{i'i} T^B_{j'j} f^{ABC} \frac{\vec{q} \cdot \vec{\varepsilon}^{\,C}}{\vec{q}^{\,4}} \ ,
\end{eqnarray}
where $q$ is the momentum running on the gluon propagator.
 
Going back to the coordinate space, we find this leads to the following interacting Hamiltonian
\begin{eqnarray}
H_{int}^{(2)} = - g_S \alpha_S T^A_{i'i} T^B_{j'j} f^{ABC} \vec{r} \cdot \vec{G}^{C} \ .
\end{eqnarray}
Here the strong coupling $\alpha_S$ arise from the coupling of $Y$ and $\bar Y$ with the Coulomb gluons, which is located at the vertex connecting the initial scattering state and the final bound state (next to the color indices ``A'' and ``B'' in the third diagram of Fig~\ref{radiation}). We decide to evaluate $\alpha_s$ at the energy scale which is the average of the inverse Bohr radius and initial $Y$ particle momentum.

Sandwiching $H_{int}^{(2)}$ between the bound and scattering states, we get a new contribution to the transition matrix element
\begin{eqnarray}\label{vfi2}
V_{fi}^{(2)} = i g_S\delta^{CD}\sqrt\frac{3}{8} \left\langle \Psi_f \left| \frac{\alpha_S}{r} \vec{r} \right| \Psi_i \right\rangle \cdot \vec{\varepsilon}^{\, C}_{\lambda} \ .
\end{eqnarray}

Summing up Eqs.~(\ref{vfi1}) and (\ref{vfi2}), we get the total transition matrix element,
\begin{eqnarray}
V_{fi} = V_{fi}^{(1)} + V_{fi}^{(2)} = \frac{\delta^{CD}}{\sqrt6} (-i g_S) \langle \Psi_f | \left( E_i - E_f - \Delta V_1(r) - \Delta V_2(r) \right) \vec{r} | \Psi_i \rangle \cdot \vec{\varepsilon}^{\, C}_{\lambda} \ ,
\end{eqnarray}
where $\Delta V_2(r) = 3 \alpha_S/(2r)$ is also given by Eq.~(\ref{dv2}). We further derive the capture cross section Eq.~(\ref{colorkramers}) based on the above matrix element.

%%%%%%%%%%%%%%%%%%%%%%%%%%%%%%%%%%%%%%%%%%%%%%%
\section{Bound States and Projection Operators}\label{app:bound}
%%%%%%%%%%%%%%%%%%%%%%%%%%%%%%%%%%%%%%%%%%%%%%%

In this appendix, we define the bound states made of scalar or fermion color particle and its anti-particle, and introduce the corresponding projection operator for calculating the $S$-wave bound state decay processes.

For a scalar color particle $Y$ (stop or sbottom), we define its plane wave state to be $|\vec{k}\rangle = \sqrt{2 \omega_k} a^\dagger_{\vec{k}} |0\rangle$. Under this convention, $\langle \vec{k}|\vec{k}'\rangle = 2 \omega_k (2\pi)^3 \delta^3(\vec{k}-\vec{k}')$. Then the bound state at rest made of $Y\bar Y$ is defined as
\begin{eqnarray}
|B\rangle = \frac{\delta_{ij}}{\sqrt{3 m_Y}} \int \frac{d^3k}{(2\pi)^3} \tilde \Psi(k)\ |\vec{k}\rangle_i \otimes |-\vec{k}\,\rangle_j \ ,
\end{eqnarray}
where $i,j$ are the color indices, and $\tilde \Psi(k)$ is the Fourier transformation of the bound state wavefunction in the momentum space, $\tilde \Psi(k) = \int d^3 x \Psi(x) e^{i \vec{k}\cdot \vec{x}}$. It is straightforward to verify that the inner product of two bound state at rest is, $\langle B | B \rangle = 4 m_Y (2\pi)^3 \delta^3(0)$, which agrees with the definition for fundamental particles.
When calculating the annihilation decay of a bound state made of $Y\bar Y$, we first calculate the amplitude for $Y\bar Y \to {\rm final}$, and then
use the $Y\bar Y$ fields to act on the above bound state. Effectively, the amplitude for bound state decay can be obtained by contracting the ($Y\bar Y \to {\rm final}$) amplitude with the following projection operator,
\begin{eqnarray}
\Pi = \frac{\delta_{ij}}{\sqrt{3 m_Y}} \Psi(0) \ .
\end{eqnarray}

Similarly, for a bound state made of colored fermion partner and its antiparticle (such as the top partner), the $\eta_T$ and $\Upsilon_T$ states are defined as
\begin{eqnarray}
|\eta_T\rangle = \frac{\delta_{ij}}{\sqrt{24 m_Y^3}} \sum_{s,s'}\int \frac{d^3k}{(2\pi)^3} \tilde \Psi(k)\  |\vec{k}, s\rangle_i \otimes |-\vec{k}, s'\rangle_j \ \bar u_s(\vec{k}) \gamma_5 v_{s'}(\vec{k}) \ , \nonumber \\
|\Upsilon_T\rangle_\lambda = \frac{\delta_{ij}}{\sqrt{24 m_Y^3}} \sum_{s,s'}\int \frac{d^3k}{(2\pi)^3} \tilde \Psi(k)\ |\vec{k}, s\rangle_i \otimes |-\vec{k}, s'\rangle_j \ \bar u_s(\vec{k}) \gamma^\mu v_{s'}(\vec{k}) \varepsilon_\mu^\lambda \ .
\end{eqnarray}
To calculate the amplitude for bound state decay in this case, we first calculate the amplitude for $Y\bar Y \to {\rm final}$ with the external $Y\bar Y$ spinors amputated, then
contract and trace over it in together with the projector operators of the form
\begin{eqnarray}
&&\Pi_\eta = \frac{\delta_{ij}}{\sqrt{24 m_Y^3}} \Psi(0) \left(\frac{\not \!P}{2} + m_Y\right) \gamma_5 \left(\frac{\not \!P}{2} - m_Y\right) \ , \nonumber \\
&&\Pi_\Upsilon = \frac{\delta_{ij}}{\sqrt{24 m_Y^3}} \Psi(0) \left(\frac{\not \!P}{2} + m_Y\right) \gamma^\mu \left(\frac{\not \!P}{2} - m_Y\right) \varepsilon_\mu^\lambda \ ,
\end{eqnarray}
where $P^\mu$ is the four momentum of the decaying bound state.

\end{document}